\documentclass[12pt, a4paper]{article}
\usepackage{a4wide,amssymb,amsmath,amscd}

\newcommand{\T}{{\mathbb T}}
\newcommand{\Z}{{\mathbb Z}}
\newcommand{\R}{{\mathbb R}}
\newcommand{\C}{{\mathbb C}}

\renewcommand{\Re}{\mathrm{Re}}
\renewcommand{\Im}{\mathrm{Im}}

\begin{document}

\topmargin -2pt

\headheight 0pt

\begin{flushright}
{\tt KIAS-P08038}
\end{flushright}


\begin{center}
{\Large \bf Noncommutative Supertori in Two Dimensions } \\

\vspace{10mm}

{\sc Ee Chang-Young}\footnote{cylee@sejong.ac.kr}\\
{\it Department of Physics, Sejong University, Seoul 143-747, Korea}\\

\vspace{5mm}

{\sc Hoil Kim}\footnote{hikim@knu.ac.kr}\\

{\it Department of Mathematics, Kyungpook National University,\\
Taegu 702-701, Korea}\\

\vspace{2mm}

and \\

\vspace{2mm}

{\sc Hiroaki Nakajima}\footnote{nakajima@skku.edu}\\

{\it Department of Physics and Institute of Basic Science, \\
Sungkyunkwan University, Suwon 440-746, Korea}\\

\vspace{10mm}

{\bf ABSTRACT} \\
\end{center}

\vspace{2mm}

\noindent
 First we consider the deformations of superspaces with $\mathcal{N}=(1,1)$
and $\mathcal{N}=(2,2)$ supersymmetries in two dimensions.
Among these the construction of noncommutative supertorus with odd spin structure
is possible only
in the case of $\mathcal{N}=(2,2)$ supersymmetry broken down to $\mathcal{N}=(1,1)$.
However, for the even spin structures the construction of noncommutative supertorus
is possible for both $\mathcal{N}=(1,1)$
and $\mathcal{N}=(2,2)$ cases.
The spin structures are realized by implementing the translational
properties along the cycles of commutative supertorus in the operator version:
Odd spin structure is realized by the translation in the fermionic direction in the same manner
as in the construction of noncommutative torus, and even spin structures are
realized with appropriate versions of the spin angular momentum operator.
\\

\setcounter{footnote}{0}



\thispagestyle{empty}

\newpage
\section{Introduction}

Ever since the work of Connes, Douglas, and Schwarz \cite{cds}
on the toroidal compactification of M(atrix) theory
using the concept of noncommutative geometry \cite{conn},
the noncommutative torus and noncommutative geometry in general
has become a household subject in string theory \cite{jp,gsw}.

Noncommutative geometry naturally appears in string
theory: Low-energy effective theory of D-branes in a
background NSNS $B$-field becomes the noncommutative field theory
where the spacetime coordinates $x^{\mu}$ are noncommutative,
$[x^{\mu}, x^{\nu}] \neq 0$ \cite{cds,SeWi,ChuHo}.
If we turn on the background RR field, the low-energy
effective theory of D-branes becomes the field theory on
non(anti)commutative superspace of which the fermionic coordinate
$\theta^{\alpha}$ has nontrivial commutation relation
$\{\theta^{\alpha}, \theta^{\beta}\} \neq 0$
\cite{Klemm:2001yu,OoVa,de Boer:2003dn,Se,BeSe}. Gauge theories on
non(anti)commutative superspace are studied extensively
\cite{ArItOh,FeIvLeSoZu,Ito:2005jj,CaIvLeQu}.

Toroidal compactification in string theory with the above mentioned background fields
then naturally leads to noncommutative supertorus.
Although the noncommutative torus is a very well known subject,
its supersymmetric version, the noncommutative supertorus, still
remains virtually unknown.
Commutative supertorus was constructed by Rabin and Freund \cite{rf1988} based on the
work of Crane and Rabin \cite{cr1988} on super Riemann surfaces.
The supertorus was obtained as the quotient of superplane by a subgroup of Osp$(1|2)$
which acts properly discontinuously on the plane together with the metrizable condition.
These two conditions boil down to proper latticing of the superplane,
and can be expressed as appropriate translation properties along the cycles of the torus.

In this paper,
adapting the guideline of defining noncommutative torus
we construct noncommutative supertorus.
Noting that the construction of bosonic noncommutative torus is guided by
the classical translation properties of the commutative torus along its cycles,
we construct the noncommutative supertorus with the following two guidelines:
Express the translations along the cycles of the supertorus in the operator language,
and implement the spin structures of supertorus for even and odd translations
with the spin angular momentum operator in appropriate representations.
Noncommutative torus is defined by embedding the lattice \cite{rief88,manin1-tr,manin3,ek08}
into the Heisenberg group \cite{mumford,sthan98,jrosenberg}.
The lattice embedding determines how the generators of noncommutative torus,
which correspond to the translation operators along the cycles of the commutative torus,
would act on the module of the noncommutative torus.
The Heisenberg group can be regarded as a
central extension of commutative space,
which is equivalent to a deformation
of space by constant noncommutativity.
Recently, we constructed the super Heisenberg group \cite{ekn08}
as a central extension of ordinary superspace,
which is equivalent to the deformation
of superspace by constant noncommutativity and nonanticommutativity.
Based on our construction of super Heisenberg group, we
define the embedding maps for noncommutative supertori in two dimensions.
Incorporating the spin structures is an additional task implementing
the translational properties along the cycles of commutative supertorus in
the operator language.
 For the odd spin structure case it is realized by the translation
 in the fermionic direction as in the construction of noncommutative torus.
 For the even spin structure cases, they are
realized with appropriate representations of the spin angular momentum operator.

This paper is organized as follows. In section 2,
we briefly recall the definition of noncommutative torus
and the projective module on which it acts.
Construction of the noncommutative tori via embedding the lattice into Heisenberg group
is also explained in the Heisenberg representation.
In section 3, we
briefly review the commutative supertori.
In section 4,
we first recall the deformation of superspace in relation with super Heisenberg group.
Then, we explicitly perform the construction of noncommutative supertorus via
embedding map in the cases of $\mathcal{N}=(1,1)$ and $\mathcal{N}=(2,2)$.
We conclude in section 5.
\\

%
\section{Noncommutative tori}

\noindent
Noncommutative torus (${\T}^d_\theta$) is an algebra
defined by generators $U_1,\cdots,U_d$ obeying the following
relations:
\begin{align}
U_iU_j=e^{2 \pi i \theta_{ij}}U_jU_i , ~~~ i,j=1,\cdots,d,
\label{nctorus}
\end{align}
where $(\theta_{ij})$ is a real $d\times d$ anti-symmetric matrix.
%
${\T}^d_\theta$ defines the involutive algebra
\begin{equation}
{\cal A}_\theta^d=
\{\sum a_{i_1\cdots i_d}U_1^{i_1}\cdots U_d^{i_d}\mid
a=(a_{i_1\cdots i_d})\in {\cal S}({\Z}^d)\},
\end{equation}
 where ${\cal
S}({\Z}^d)$ is the Schwartz space of sequences with rapid decay.

Every projective module over a smooth algebra ${\cal
A}^{d}_{\theta}$ can be represented by a direct sum of modules of
the form ${\cal S}({\R}^p\times{\Z}^q\times F)$ \cite{rief88}, the linear space
of Schwartz functions on ${\R}^p\times{\Z}^q\times F$, where
$2p+q=d$ and $F$ is a finite abelian group.
Let $D$ be a lattice in ${\cal G}=M\times \widehat{M}$, where
$M={\R}^p\times{\Z}^q\times F$ and $\widehat{M}$ is its dual.
The embedding map $\Phi$ under which $D$ is the image of ${\Z}^d$
 determines a projective module $E$ on which the algebra of the noncommutative torus acts.

In the Heisenberg representation the operators $U$'s are
 defined by
\begin{equation}
U_{(m,\hat s)}f(r)=e^{2\pi i <r, \hat s>}f(r+m), ~~ m,r \in M, ~ \hat s \in \widehat{M}, ~
  f \in E ,
\label{heisrep}
\end{equation}
where $<r, \hat s>$ is a usual inner product between $M$ and $\widehat{M}$.
Here, the vector $(m,\hat s)$ can be mapped into an element of the Heisenberg group
which we explain next.

The Heisenberg group, $\mathit{Heis}({\R}^{2n},\psi)$, is defined as
follows.
For $t, t' \in U(1)$, and $(x,y),(x',y') \in \R^{2n}$, we define
the product for $(t,x,y),(t',x',y')  \in \mathit{Heis}({\R}^{2n},\psi)$,
\begin{align}
(t,x,y) \cdot (t',x',y') = (t+t' + \psi(x,y;x',y'), x+x',y+y'),
\label{multip}
\end{align}
where $  \psi:~~ {\R}^{2n} \times {\R}^{2n} \longrightarrow \R,$ satisfies
the cocycle condition
\begin{equation}
\psi(x,y;x',y')\psi(x+x',y+y';x'',y'')=
\psi(x,y;x'+x'',y'+y'')\psi(x',y';x'',y''),
\end{equation}
which is a necessary and sufficient condition for the multiplication
to be associative.
There is an exact sequence
\begin{equation}
0 \rightarrow \mathbb{R} \mathop{\rightarrow}^{i}
\mathit{Heis}(\mathbb{R}^{2n},\psi) \mathop{\rightarrow}^{j}
\mathbb{R}^{2n} \rightarrow 0
\end{equation}
called a central extension, with the inclusion $i(t)=(t,0,0)$ and the projection $j(t,x,y)=(x,y)$,
where $i(\mathbb{R})$ is the center in
$\mathit{Heis}({\R}^{2n},\psi)$.
The previously appeared vector  $(m,\hat s)$ in (\ref{heisrep}) corresponds to
a vector $(x,y) \in {\R}^{2n}$ in the above description of the Heisenberg group.

Now, we consider an explicit form of the embedding in a typical case
where $M$ in (\ref{heisrep}) is given by $M={\R}^{p}$.
In this case, one can define the embedding map in the canonical form as
\begin{equation}
\Phi=\begin{pmatrix}\Theta &0 \\
                0&   I \end{pmatrix} : = (x_{i,j}), ~~
{\rm where} ~~
\Theta = {\rm diag}(\theta_1, \cdots, \theta_p), \ \   i,j = 1, \cdots, 2p,
\label{embmap}
\end{equation}
then the Heisenberg representation is given as follows.
\begin{equation}
\label{Heis_can}
 (U_j f)(s_1, \cdots, s_p)  : =  (U_{\vec{e}_j}f)(\vec{s}) =  \exp(2\pi i \sum_{k=1}^p s_k
x_{k+p,j}  )f(\vec{s} +\vec{x}_j),
~~ {\rm for} \ ~ \ j=1, \cdots, 2p ,
\end{equation}
where  $~ \vec{e}_j=(x_{1,j},\dots, x_{2p,j})$.
The above can be redisplayed as
\begin{equation}
(U_j f)(\vec{s}) = f(\vec{s}  + \vec{\theta}) , ~~
 (U_{j+p} f)(\vec{s})=e^{2\pi i s_j } f(\vec{s}), ~~~ j,k=1, \cdots, p,
 \end{equation}
with $\vec{s}=(s_1, \cdots, s_p), ~ \vec{x}_j=(x_{1,j}, \cdots,
x_{p,j})$ and $ \vec{s}, \vec{x}_j \in {\R}^p $.
 Here, $U_j$'s satisfy
\begin{equation}
U_j U_{j+p}  =e^{2\pi i \theta_j} U_{j+p}U_j, ~~ {\rm otherwise}~~
U_j U_k = U_k U_j .
\end{equation}
In the general embedding case,
we will use the Manin's representation \cite{manin3} in
which (\ref{Heis_can}) becomes
\begin{equation}
\label{Heis_manin}
(U_{\vec{e}_j}f)(\vec{s}) :=  \exp(2\pi i \sum_{k=1}^p s_k
x_{k+p,j}+ \frac{1}{2}\sum_{k=1}^p x_{k,j}
x_{k+p,j}  )f(\vec{s} +\vec{x}_j),
~~ {\rm for} \ ~ \ j=1, \cdots, 2p .
\end{equation}
This representation corresponds to the second representation of the
Heisenberg group in \cite{ekn08}.
\\

%

%


\section{Commutative supertori}\label{sptori}
A two-dimensional commutative torus is given by
$\T^{2}=\C / (\Z + \tau \Z)$, where $\tau$ is a complex structure.
In a similar way, as the result of
the supersymmetric version of the uniformization theorem \cite{cr1988},
a supertorus is given by
a quotient of the two-dimensional superspace by a subgroup of
Osp$(\mathcal{N}|2)$ 
which is the anomaly-free part of the superconformal group
\cite{cr1988,rf1988}.
The action of this subgroup gives the cycles of the supertorus.
When we consider $\mathcal{N}=(1,1)$ superspace
spanned by supercoordinates $(z,\bar{z},\theta,\bar{\theta})$
as an ambient space, the action of Osp$(1|2)$ 
in the holomorphic sector is given by
\begin{align}
z&\to z'=\frac{az+b}{cz+d}+\theta\frac{\gamma z+\delta}{(cz+d)^{2}},\quad
ad-bc=1,\notag\\[2mm]
\theta&\to\theta'=\frac{\gamma z+\delta}{cz+d}+\frac{\theta}{cz+d}
\left(1+\frac{1}{2}\delta\gamma\right).
\label{sc}
\end{align}
The condition for the subgroup is i) the supertorus is metrizable,
\textit{i.e.} the metric of the supertorus is invariant under
this subgroup, ii) this subgroup acts properly discontinuously,
which means that $z$ and $z'$ have disjointed neighborhoods. From
these conditions, we have $c=\gamma=0$ and \eqref{sc} is reduced to
\begin{align}
z&\to z'=a^{2}z+ab+a^{2}\theta\delta,\notag\\
\theta&\to\theta'=a(\theta+\delta),
\label{sg}
\end{align}
where $a=\pm 1$. The sign of $a$ in each cycle determines the spin structure.
The action of the subgroup
\eqref{sg} is more simplified by the similarity transformation via an element
of Osp$(1|2)$. 
Then it is classified to the four cases according to
the spin structure of the supertorus as follows:
\begin{itemize}
\item $(+,+)$ structure
\begin{align}
(z,\,\theta)\to (z+1,\,\theta),\quad
(z,\,\theta)\to (z+\tau+\theta\delta,\,\theta+\delta).
\label{odd}
\end{align}
\item $(+,-)$ structure
\begin{align}
(z,\,\theta)\to (z+1,\,\theta),\quad
(z,\,\theta)\to (z+\tau,\,-\theta).
\label{even1}
\end{align}
\item $(-,+)$ structure
\begin{align}
(z,\,\theta)\to (z+1,\,-\theta),\quad
(z,\,\theta)\to (z+\tau,\,\theta).
\label{even2}
\end{align}
\item $(-,-)$ structure
\begin{align}
(z,\,\theta)\to (z+1,\,-\theta),\quad
(z,\,\theta)\to (z+\tau,\,-\theta).
\label{even3}
\end{align}
\end{itemize}
Here $\tau$ and $\delta$ are the moduli parameters of the supertorus.
The action of the subgroup \eqref{odd}--\eqref{even3}
for the anti-holomorphic sector $(\bar{z},\,\bar{\theta})$ is obtained by
complex conjugation.
The $(+,+)$ structure is also called the odd spin structure
and the other three structures are called the even spin structures.

$\mathcal{N}=(2,2)$ supertorus can be also constructed in a similar manner.
Let $(z,\bar{z},\theta^{\pm},\bar{\theta}^{\pm})$ be supercoordinates
of $\mathcal{N}=(2,2)$ superspace. Then the cycles in
$\mathcal{N}=(2,2)$ supertorus with the odd spin structure are given by
\begin{align}
(z,\,\theta^{+},\,\bar{\theta}^{+})&\to
(z+1,\,\theta^{+},\,\bar{\theta}^{+}),\notag\\
(z,\,\theta^{+},\,\bar{\theta}^{+})&\to
(z+\tau+\bar{\theta}^{+}\delta^{+}+\theta^{+}\bar{\delta}^{+},\,
\theta^{+}+\delta^{+},\,
\bar{\theta}^{+}+\bar{\delta}^{+}).
\end{align}
In the case of the $(+,-)$ structure, the cycles in the supertorus are given by
\begin{align}
(z,\,\theta^{+},\,\bar{\theta}^{+})\to
(z+1,\,\theta^{+},\,\bar{\theta}^{+}),\quad
(z,\,\theta^{+},\,\bar{\theta}^{+})\to
(z+\tau,\,-\theta^{+},\,-\bar{\theta}^{+}).
\end{align}
The cycles in the other even spin structures are also obtained similarly.
\\

%

\section{Noncommutative supertori}


Noncommutative supertorus is defined by embedding the lattice to
the super Heisenberg group. The super Heisenberg group is given by the
central extension of ordinary superspace,
which is equivalent to the deformation
of superspace by constant noncommutativity and nonanticommutativity.

\subsection{Deformation of superspace}
First we consider the deformation of $\mathcal{N}=(1,1)$ superspace
spanned by supercoordinates $(X^{1}, X^{2}, \theta, \bar{\theta})$.
Supercharges and supercovariant derivatives are defined by
\begin{align}
Q&=\frac{\partial}{\partial\theta}-\theta\frac{\partial}{\partial Z},&
\bar{Q}&=\frac{\partial}{\partial\bar{\theta}}
-\bar{\theta}\frac{\partial}{\partial \bar{Z}},
\\
D&=\frac{\partial}{\partial\theta}+\theta\frac{\partial}{\partial Z},&
\bar{D}&=\frac{\partial}{\partial\bar{\theta}}
+\bar{\theta}\frac{\partial}{\partial \bar{Z}},
\end{align}
where $Z$, $\bar{Z}$ are the complex coordinates given by
\begin{align}
Z&=X^{1}+iX^{2}, & \bar{Z}&=X^{1}-iX^{2},\\
\frac{\partial}{\partial Z}&=
\frac{1}{2}
\left(\frac{\partial}{\partial X^{1}}-i\frac{\partial}{\partial X^{2}}\right),
&
\frac{\partial}{\partial \bar{Z}}&=
\frac{1}{2}
\left(\frac{\partial}{\partial X^{1}}+i\frac{\partial}{\partial X^{2}}\right).
\end{align}
Now we try to introduce the deformation for fermionic coordinates
 keeping the reality condition and preserving
full or partial supersymmetry.
However it turns out that it is impossible to perform the above deformation
preserving the Heisenberg group structure.
For instance, consider the Moyal product of the form,
\begin{equation}
\ast'=\exp\left[\frac{i}{2}\varTheta\epsilon^{\mu\nu}
\overleftarrow{\frac{\partial}{\partial X^{\mu}}}
\overrightarrow{\frac{\partial}{\partial X^{\nu}}}
-\frac{C}{2}\left(\overleftarrow{Q}\overrightarrow{\bar{Q}}
+\overleftarrow{\bar{Q}}\overrightarrow{Q}\right)
\right].
\label{q1}
\end{equation}
Then supersymmetry is completely broken,
since under \eqref{q1} the Leibniz rule
\begin{equation}
Q(f\ast'g)=(Qf)\ast'g+(-1)^{|f|}f\ast'(Qg),\quad
\bar{Q}(f\ast'g)=(\bar{Q}f)\ast'g+(-1)^{|f|}f\ast'(\bar{Q}g)
\label{leibniz}
\end{equation}
is not satisfied, where $|f|$ is zero for Grassmann even $f$ and unity for
Grassmann odd $f$.
The next candidate of the Moyal product is
\begin{equation}
\ast''=\exp\left[\frac{i}{2}\varTheta\epsilon^{\mu\nu}
\overleftarrow{\frac{\partial}{\partial X^{\mu}}}
\overrightarrow{\frac{\partial}{\partial X^{\nu}}}
-\frac{C'}{2}\left(\overleftarrow{D}\overrightarrow{\bar{D}}
+\overleftarrow{\bar{D}}\overrightarrow{D}\right)
\right].
\end{equation}
In this case supersymmetry is preserved since \eqref{leibniz} holds
but the corresponding algebra in the operator formalism becomes
\begin{align}
[X^{1},X^{2}]&=i\left[\varTheta+\frac{C'}{4}
(\bar{\theta}\theta-\theta\bar{\theta})\right],&
[X^{1},\theta]&=-\frac{C'}{2}\bar{\theta},&
[X^{1},\bar{\theta}]&=-\frac{C'}{2}\theta,
\notag\\
[X^{2},\theta]&=-\frac{iC'}{2}\bar{\theta},&
[X^{2},\bar{\theta}]&=-\frac{iC'}{2}\theta,&
\{\theta,\bar{\theta}\}&=C',
\end{align}
which is not a super Heisenberg algebra since the commutator and anticommutator do not belong to the center.
In order to obtain a super Heisenberg algebra,
the parameter $C'$ should vanish. Thus we consider the case of $C'=0$
and use the following Moyal product
\begin{equation}
\ast=\exp\left(\frac{i}{2}\varTheta\epsilon^{\mu\nu}
\overleftarrow{\frac{\partial}{\partial X^{\mu}}}
\overrightarrow{\frac{\partial}{\partial X^{\nu}}}\right).
\end{equation}
The corresponding algebra becomes
\begin{equation}
[X^{1},X^{2}]=i\varTheta, \quad \text{others}=0, \label{def11}
\end{equation}
which can be represented as the action on the module $f(s,\theta,\bar{\theta})$ as follows.
\begin{equation}
X^{1}f(s,\theta,\bar{\theta})=i\frac{\partial}{\partial s}
f(s,\theta,\bar{\theta}),\quad
X^{2}f(s,\theta,\bar{\theta})=sf(s,\theta,\bar{\theta}),
\label{12}
\end{equation}
where we set $\varTheta=1$ by change of normalization of the operators.
\\

Next we consider $\mathcal{N}=(2,2)$ superspace spanned by supercoordinates
$(X^{1}, X^{2}, \theta^{\pm}, \bar{\theta}^{\pm})$.
Supercharges and supercovariant derivatives on $\mathcal{N}=(2,2)$ superspace
are defined by
\begin{align}
Q_{+}&=\frac{\partial}{\partial\theta^{+}}
-\bar{\theta}^{+}\frac{\partial}{\partial Z},&
\bar{Q}_{+}&=\frac{\partial}{\partial\bar{\theta}^{+}}
-\theta^{+}\frac{\partial}{\partial Z},
\\
D_{+}&=\frac{\partial}{\partial\theta^{+}}
+\bar{\theta}^{+}\frac{\partial}{\partial Z},&
\bar{D}_{+}&=\frac{\partial}{\partial\bar{\theta}^{+}}
+\theta^{+}\frac{\partial}{\partial Z},
\end{align}
and $(Q_{-}, \bar{Q}_{-}, D_{-}, \bar{D}_{-})$ which are obtained by
the complex conjugation $Z\leftrightarrow\bar{Z}$,
$\theta^{+}\leftrightarrow\theta^{-}$,
$\bar{\theta}^{+}\leftrightarrow\bar{\theta}^{-}$.
The deformation of $\mathcal{N}=(2,2)$ superspace which has the structure of
the super Heisenberg algebra is given by the following two types of deformations
\footnote{There is another deformation which corresponds to the case of
$C_{Q},\,C_{D}=0$. However this deformation gives the similar result with
$\mathcal{N}=(1,1)$ supertorus. Thus we consider the case of
$C_{Q},\,C_{D}\neq 0$.}:
\begin{align}
\ast_{Q}&=\exp\left[\frac{i}{2}\varTheta\epsilon^{\mu\nu}
\overleftarrow{\frac{\partial}{\partial X^{\mu}}}
\overrightarrow{\frac{\partial}{\partial X^{\nu}}}
-\frac{C_{Q}}{2}\left(\overleftarrow{Q_{+}}\overrightarrow{Q_{-}}
+\overleftarrow{Q_{-}}\overrightarrow{Q_{+}}\right)
\right],\label{q-def}\\
\ast_{D}&=\exp\left[\frac{i}{2}\varTheta\epsilon^{\mu\nu}
\overleftarrow{\frac{\partial}{\partial X^{\mu}}}
\overrightarrow{\frac{\partial}{\partial X^{\nu}}}
-\frac{C_{D}}{2}\left(\overleftarrow{D_{+}}\overrightarrow{D_{-}}
+\overleftarrow{D_{-}}\overrightarrow{D_{+}}\right)
\right].\label{d-def}
\end{align}
The deformations \eqref{q-def} and \eqref{d-def} are called $Q$-deformation \cite{Se} and
$D$-deformation \cite{Klemm:2001yu}, respectively.
The nontrivial commutation relation in the operator formalism
with $Q$-deformation becomes
\begin{align}
[X^{1}, X^{2}]&=i\varTheta-\frac{i}{2}C_{Q}\bar{\theta}^{+}\bar{\theta}^{-},&
[X^{1}, \theta^{+}]&=\frac{1}{2}C_{Q}\bar{\theta}^{-},&
[X^{1}, \theta^{-}]&=\frac{1}{2}C_{Q}\bar{\theta}^{+},
\notag\\
[X^{2}, \theta^{+}]&=\frac{i}{2}C_{Q}\bar{\theta}^{-},&
[X^{2}, \theta^{-}]&=-\frac{i}{2}C_{Q}\bar{\theta}^{+},&
\{\theta^{+}, \theta^{-}\}&=C_{Q}.\label{crq}
\end{align}
In the case of $D$-deformation, it becomes
\begin{align}
[X^{1}, X^{2}]&=i\varTheta-\frac{i}{2}iC_{D}\bar{\theta}^{+}\bar{\theta}^{-},&
[X^{1}, \theta^{+}]&=-\frac{1}{2}C_{D}\bar{\theta}^{-},&
[X^{1}, \theta^{-}]&=-\frac{1}{2}C_{D}\bar{\theta}^{+},
\notag\\
[X^{2}, \theta^{+}]&=-\frac{i}{2}C_{D}\bar{\theta}^{-},&
[X^{2}, \theta^{-}]&=\frac{i}{2}C_{D}\bar{\theta}^{+},&
\{\theta^{+}, \theta^{-}\}&=C_{D}.\label{crd}
\end{align}
The algebras \eqref{crq} and \eqref{crd} can be represented on
the module $f(s,\eta,\bar{\theta}^{+},\bar{\theta}^{-})$ as
\begin{align}
X^{1}f(s,\eta,\bar{\theta}^{+},\bar{\theta}^{-})&=
\left(i\frac{\partial}{\partial s}
\pm\frac{1}{2}\bar{\theta}^{+}\frac{\partial}{\partial\eta}
\pm\frac{1}{2}\bar{\theta}^{-}\eta
\right)f(s,\eta,\bar{\theta}^{+},\bar{\theta}^{-}),
\notag\\[2mm]
X^{2}f(s,\eta,\bar{\theta}^{+},\bar{\theta}^{-})&=\left(s
\mp\frac{i}{2}\bar{\theta}^{+}\frac{\partial}{\partial\eta}
\pm\frac{i}{2}\bar{\theta}^{-}\eta
\right)f(s,\eta,\bar{\theta}^{+},\bar{\theta}^{-}),
\notag\\[2mm]
\theta^{+}f(s,\eta,\bar{\theta}^{+},\bar{\theta}^{-})&=
\frac{\partial}{\partial\eta}f(s,\eta,\bar{\theta}^{+},\bar{\theta}^{-}),
\notag\\[2mm]
\theta^{-}f(s,\eta,\bar{\theta}^{+},\bar{\theta}^{-})&=
\eta f(s,\eta,\bar{\theta}^{+},\bar{\theta}^{-}),
\label{n2alg}
\end{align}
where the double signs correspond to $Q$- and $D$-deformation respectively and
we set $\varTheta=C_{Q}=C_{D}=1$.
Note that in the both cases $\bar{\theta}^{+}$ and $\bar{\theta}^{-}$ belong to
the center and the resulting algebra are the super Heisenberg algebra.
Supersymmetry is broken to $\mathcal{N}=(1,1)$ in $Q$-deformation
\cite{Se}, but it is unbroken in $D$-deformation
\cite{Klemm:2001yu}.

\subsection{Noncommutative supertori}

 First, we consider the construction of noncommutative(NC) supertorus
for $\mathcal{N}=(1,1)$ supersymmetry.
In order to construct the NC supertorus, we have to get the operators
which generate the translation along the cycles of the NC supertorus.
However, we do not know how to construct the equations corresponding to
\eqref{odd}--\eqref{even3} in NC supertorus since we do not have
geometrical notion in noncommutative (super)space
such as the metric and the disjointed neighborhoods.
Thus we construct the generators by simply mimicking the commutative case.
In the odd spin structure, The generators of the NC supertorus
satisfy
\begin{align}
UX^{1}U^{-1}&=X^{1}+1,& UX^{2}U^{-1}&=X^{2},\notag\\
U\theta U^{-1}&=\theta,& U\bar{\theta} U^{-1}&=\bar{\theta},
\notag\\[3mm]
VX^{1}V^{-1}&=X^{1}+\Re(\tau+\theta\delta),&
VX^{2}V^{-1}&=X^{2}+\Im(\tau+\theta\delta),
\notag\\
V\theta V^{-1}&=\theta+\delta,& V\bar{\theta} V^{-1}&=
\bar{\theta}+\bar{\delta},
\label{trans11_odd}
\end{align}
which corresponds to \eqref{odd}.
Here the supercoordinates $(X^{1}, X^{2}, \theta, \bar{\theta})$
satisfy the commutation relation \eqref{def11}.
Then the explicit form of the generators $U,\,V$ is given by
\begin{align}
U&=\exp(is),\notag\\
V&=\exp\left[i(\Re\,\tau)s+(\Im\,\tau)\frac{\partial}{\partial s}
+\delta\mathcal{Q}+\bar{\delta}\bar{\mathcal{Q}}\right]\notag\\
&=\exp\left[i\Re(\tau+\theta\delta)s+\Im(\tau+\theta\delta)
\frac{\partial}{\partial s}
+\delta\frac{\partial}{\partial\theta}
+\bar{\delta}\frac{\partial}{\partial\bar{\theta}}\right],
\label{n1uv}
\end{align}
where
$\mathcal{Q},\,\bar{\mathcal{Q}}$ are the representation of
the supercharges $Q,\,\bar{Q}$ on the module and are defined by
\begin{equation}
\label{qq11}
\mathcal{Q}=\frac{\partial}{\partial\theta}
-\frac{i}{2}\theta\left(s-\frac{\partial}{\partial s}\right),\quad
\bar{\mathcal{Q}}=\frac{\partial}{\partial\bar{\theta}}
-\frac{i}{2}\bar{\theta}\left(s+\frac{\partial}{\partial s}\right).
\end{equation}
Here an important comment is in order.
Although the generators $U,\,V$ correctly give the translational property
along the cycles, \eqref{trans11_odd}, the generator $V$ does not belong
to a representation of the super Heisenberg group which is the prerequisite
for noncommutative supertorus. In the super Heisenberg group representation of
noncommutative superspace, the coordinates play the role of the generators
and their (anti)commutators should be constant. However, here
$Q,\ \bar{Q}$ plays the role of $\theta, \bar{\theta}$, and their
anticommutator is not constant, which can be seen in \eqref{qq11}.

This can be also seen from the embedding map picture.
 From \eqref{n1uv}, the embedding map can be written as
\begin{equation}
\Phi=\,\bordermatrix{&U&V\cr s&1&\Re(\tau+\theta\delta)\cr \theta&0&0\cr
\bar{\theta}&0&0\cr \frac{\partial}{\partial s}&0&\Im(\tau+\theta\delta)\cr
\frac{\partial}{\partial\theta}&0&\delta\cr
\frac{\partial}{\partial\bar{\theta}}&0&\bar{\delta}\cr
}.
\end{equation}
In the embedding map for $V$, we see that the coordinate variable $\theta$ appears.
This makes the action of the operator $V$ a lot different from the allowed
one given by \eqref{Heis_manin}.
This is in turn reflected in the commutation relation between $U$ and $V$,
\begin{equation}
\label{odd11}
UV=\exp[-i\,\Im(\tau+\theta\delta)]VU .
\end{equation}
We now see that it does not satisfy the defining relation for noncommutative
torus since $\theta_{ij}$ in \eqref{nctorus} fails to be a constant due to
the the presence of the coordinate variable $\theta$ in \eqref{odd11}.
\\

 For the even spin structures, \eqref{even1}--\eqref{even3} contain
the sign change $\theta\to -\theta$, $\bar{\theta}\to -\bar{\theta}$.
Hence in order to construct the generators, we need the generator of
sign change for $\theta$ and $\bar{\theta}$, which is given by
the spin angular momentum operator
\begin{equation}
J=\frac{1}{2}\left(\theta\frac{\partial}{\partial\theta}
-\bar{\theta}\frac{\partial}{\partial\bar{\theta}}\right).
\end{equation}
Then the generators of the NC supertorus with even spin structures are
\begin{itemize}
\item $(+,-)$ structure
\begin{align}
U&=\exp(is),\notag\\
V&=\exp(2\pi iJ)
\exp\left[i(\Re\,\tau)s+(\Im\,\tau)\frac{\partial}{\partial s}\right].
\end{align}
\item $(-,+)$ structure
\begin{align}
U&=\exp(2\pi iJ)\exp(is),\notag\\
V&=\exp\left[i(\Re\,\tau)s+(\Im\,\tau)\frac{\partial}{\partial s}\right].
\end{align}
\item $(-,-)$ structure
\begin{align}
U&=\exp(2\pi iJ)\exp(is),\notag\\
V&=\exp(2\pi iJ)
\exp\left[i(\Re\,\tau)s+(\Im\,\tau)\frac{\partial}{\partial s}\right].
\end{align}
\end{itemize}
In these three structures, the commutation relation between $U$ and $V$ is
given by
\begin{equation}
UV=\exp(-i\,\Im\,\tau)VU.
\label{n1comeven}
\end{equation}
Note that in the above three cases of even spin structures no obstruction
appears in the construction of noncommutative supertorus.
\\

Next we consider the $\mathcal{N}=(2,2)$ case.
 For $\mathcal{N}=(2,2)$ supersymmetry, we proceed in the same manner as in
 the  $\mathcal{N}=(1,1)$ case.
 For the odd spin structure, the generators $U,\,V$ satisfy
\begin{align}
UX^{\mu}U^{-1}&=X^{\mu}+e_{U}^{\mu},&
U\theta^{\pm} U^{-1}&=\theta^{\pm},&
U\bar{\theta}^{\pm} U^{-1}&=\bar{\theta}^{\pm},
\notag\\
VX^{\mu}V^{-1}&=X^{\mu}+e_{V}^{\mu},&
V\theta^{\pm} V^{-1}&=\theta^{\pm}+\delta^{\pm},&
V\bar{\theta}^{\pm} V^{-1}&=\bar{\theta}^{\pm}+\bar{\delta}^{\pm},
\end{align}
where the supercoordinates satisfy the algebra \eqref{n2alg}
and the lattice vectors $e_{U}^{\mu}$ and $e_{V}^{\mu}$ are given by
\begin{align}
e_{U}^{\mu}&={}^{t}(1,0),\notag\\
e_{V}^{\mu}&={}^{t}
\bigl(\Re(\tau+\bar{\theta}^{+}\delta^{+}+\theta^{+}\bar{\delta}^{+}),\,
\Im(\tau+\bar{\theta}^{+}\delta^{+}+\theta^{+}\bar{\delta}^{+})\bigr).
\end{align}
Then the explicit form of $U,\,V$ in the $Q$-deformation \eqref{q-def} is
given by
\begin{align}
U&=\exp(is),
\notag\\
V&=\exp\biggl[i(\Re\,\tau)s+(\Im\,\tau)\frac{\partial}{\partial s}
+\delta^{+}\mathcal{Q}_{+}+\delta^{-}\mathcal{Q}_{-}\biggr]
\notag\\
&=\exp\biggl[i(\Re\,\tau)s+(\Im\,\tau)\frac{\partial}{\partial s}
+\delta^{+}\eta+\delta^{-}\frac{\partial}{\partial\eta}\biggr],
\label{n2genq}
\end{align}
where the operators
$\mathcal{Q}_{\pm}$ are the representation of
the supercharges $Q_{\pm}$ on the module, and can be simply represented by
\begin{equation}
\mathcal{Q}_{+}=\eta\cdot{}\,,\quad \mathcal{Q}_{-}=
\frac{\partial}{\partial\eta}.
\end{equation}
In the above we set $\bar{\delta}^{\pm}=0$, since
the representation of $\bar{Q}^{\pm}$ on the module is given by
\begin{equation}
\bar{\mathcal{Q}}_{+}=\frac{\partial}{\partial\bar{\theta}^{+}}
-i\eta\left(s-\frac{\partial}{\partial s}\right),\quad
\bar{\mathcal{Q}}_{-}=\frac{\partial}{\partial\bar{\theta}^{-}}
-i\frac{\partial}{\partial\eta}\left(s-\frac{\partial}{\partial s}\right),
\end{equation}
which contains the second order derivative%
\footnote{This corresponds to that the supersymmetry generated by
$\bar{Q}^{\pm}$ is broken by the $Q$-deformation.},
and thus it cannot be included in $V$.
The commutation relation between $U$ and $V$ is
given by
\begin{equation}
UV=\exp(-i\,\Im\,\tau)VU,
\end{equation}
and shows no obstruction for noncommutative supertorus.
On the other hand, in the $D$-deformation, $U$ and $V$ are obtained as
\begin{align}
U&=\exp(is),
\notag\\
V&=\exp\biggl[i(\Re\,\tau)s+(\Im\,\tau)\frac{\partial}{\partial s}
+\delta^{+}\mathcal{Q}'_{+}+\delta^{-}\mathcal{Q}'_{-}
+\bar{\delta}^{+}\bar{\mathcal{Q}}'_{+}
+\bar{\delta}^{-}\bar{\mathcal{Q}}'_{-}\biggr]
\notag\\
&=\exp\biggl[i\Re(\tau+2\bar{\theta}^{+}\delta^{+})s
+\Im(\tau+2\bar{\theta}^{+}\delta^{+})\frac{\partial}{\partial s}
\notag\\
&\qquad\qquad\qquad
{}+\delta^{+}\eta+\delta^{-}\frac{\partial}{\partial\eta}
+\bar{\delta}^{+}\frac{\partial}{\partial\bar{\theta}^{+}}
+\bar{\delta}^{-}\frac{\partial}{\partial\bar{\theta}^{-}}\biggr],
\label{n2gend}
\end{align}
where the operators
$\mathcal{Q}'_{\pm},\,\bar{\mathcal{Q}}'_{\pm}$ are the representation of
the supercharges $Q_{\pm},\,\bar{Q}_{\pm}$ on the module and are given by
\begin{align}
\mathcal{Q}'_{+}&=\eta
-i\bar{\theta}^{+}\left(s-\frac{\partial}{\partial s}\right),&
\mathcal{Q}'_{-}&=\frac{\partial}{\partial\eta}
-i\bar{\theta}^{-}\left(s+\frac{\partial}{\partial s}\right),
\notag\\
\bar{\mathcal{Q}}'_{+}&=\frac{\partial}{\partial\bar{\theta}^{+}},&
\bar{\mathcal{Q}}'_{-}&=\frac{\partial}{\partial\bar{\theta}^{-}}.
\end{align}
Then the commutation relation between $U$ and $V$ is
given by
\begin{equation}
UV=\exp[-i\,\Im(\tau+2\bar{\theta}^{+}\delta^{+})]VU.
\end{equation}
Again in this case, we see the obstruction that appeared in the
 $\mathcal{N}=(1,1)$ case.
 The $V$ in \eqref{n2gend} does not belong to the super Heisenberg group,
 thus cannot be a generator for noncommutative supertorus.

The embedding maps for both $Q$- and $D$-deformations can be written from
\eqref{n2genq} and \eqref{n2gend} as
\begin{equation}
\Phi_{Q}=\,\bordermatrix{&U&V\cr s&1&\Re\,\tau\cr
\eta&0&\delta^{+}\cr
\bar{\theta}^{+}&0&0\cr
\bar{\theta}^{-}&0&0\cr
\frac{\partial}{\partial s}&0&\Im\,\tau\cr
\frac{\partial}{\partial \eta}&0&\delta^{-}\cr
\frac{\partial}{\partial\bar{\theta}^{+}}&0&0\cr
\frac{\partial}{\partial\bar{\theta}^{-}}&0&0\cr
}\,,\qquad
\Phi_{D}=\,\bordermatrix{&U&V\cr s&1&\Re(\tau+2\bar{\theta}^{+}\delta^{+})\cr
\eta&0&\delta^{+}\cr
\bar{\theta}^{+}&0&0\cr
\bar{\theta}^{-}&0&0\cr
\frac{\partial}{\partial s}&0&\Im(\tau+2\bar{\theta}^{+}\delta^{+})\cr
\frac{\partial}{\partial \eta}&0&\delta^{-}\cr
\frac{\partial}{\partial\bar{\theta}^{+}}&0&\bar{\delta}^{+}\cr
\frac{\partial}{\partial\bar{\theta}^{-}}&0&\bar{\delta}^{-}\cr
},
\end{equation}
and the embedding map for the $D$-deformation case shows the same problem
in the  $\mathcal{N}=(1,1)$ case.
Therefore, the construction of noncommutative supertorus with odd spin structure
in two dimensions is allowed only in the $Q$-deformation with  $\mathcal{N}=(2,2)$
supersymmetry.
\\

 In the even spin structure cases, there is no obstruction for noncommutative supertorus
 as in the $\mathcal{N}=(1,1)$ case.
To see this, we need the spin operators $J_{\theta}$ and $J_{\bar{\theta}}$
of $\theta^{\pm}$ and $\bar{\theta}^{\pm}$, respectively,
as in the $\mathcal{N}=(1,1)$ case.
The explicit form of $J_{\theta}$ and $J_{\bar{\theta}}$ are given by
\begin{equation}
J_{\theta}=-\frac{1}{2}\theta^{-}\theta^{+}
=-\frac{1}{2}\eta\frac{\partial}{\partial\eta},\quad
J_{\bar{\theta}}=\frac{1}{2}\biggl(
\bar{\theta}^{+}\frac{\partial}{\partial\bar{\theta}^{+}}
-\bar{\theta}^{-}\frac{\partial}{\partial\bar{\theta}^{-}}
\biggr).
\end{equation}
Then the generators $U$ and $V$ in $(+,-)$ structure
for both $Q$- and $D$-deformations are obtained as
\begin{align}
U&=\exp(is),\notag\\
V&=\exp\bigl[2\pi i(J_{\theta}+J_{\bar{\theta}})\bigr]
\exp\left[i(\Re\,\tau)s+(\Im\,\tau)\frac{\partial}{\partial s}\right].
\end{align}
The generators in the other spin structures can be
obtained similarly. The commutation relation between $U$ and $V$ takes
the same form with \eqref{n1comeven}.
Note that
the spin structures of $\theta^{\pm}$ and $\bar{\theta}^{\pm}$
should be the same in order to preserve the algebra
\eqref{crq} and \eqref{crd}.\\

\section{Conclusion}

In this paper, we construct noncommutative supertori in two dimensions
with $\mathcal{N}=(1,1)$ and $\mathcal{N}=(2,2)$ supersymmetries.
In the $\mathcal{N}=(1,1)$ case, only the deformation of the bosonic part is allowed
to maintain the supersymmetry and the super Heisenberg group structure. 
In this case, the super Heisenberg group structure,
which is essential to be a noncommutative supertorus, is only maintained
 for the even spin structures.
In the $\mathcal{N}=(2,2)$ case, both bosonic and fermionic parts are deformed
with two types of deformations, $Q$- and $D$-deformation.
 For the odd spin structure, the generators belong to the super Heisenberg group
 only in the $Q$-deformation case, in which the supersymmetry is broken down to $\mathcal{N}=(1,1)$.
On the other hand, the super Heisenberg group structure is maintained
 in  both $Q$- and $D$-deformations for the even spin structures.
The result shows that for the odd spin structure noncommutative supertorus in two dimensions
is allowed only in the $Q$-deformation with $\mathcal{N}=(2,2)$ supersymmetry, 
while there is no obstruction for the even spin structures.
One might understand this result from the underlying property of spin structures:
Odd spin structure is related with a translation in the fermionic direction,
thus the allowed deformation which is consistent with the super Heisenberg group structure
is restricted.
Even spin structures are related with translations in the bosonic cycles only,
thus they do not interfere with the super Heisenberg group structure.

Next, we comment other aspects of our work.
The commutative supertorus was constructed by taking the quotient of the superplane
with a proper lattice,
and has certain translation properties along its cycles.
Noncommutative torus was constructed such that it maintains the properties
under the translations along the cycles of commutative torus.
We thus construct the generators of the noncommutative supertorus
such that the properties under translations along the cycles of the commutative supertorus
are maintained.

Noncommutative torus was defined by embedding the lattice
into the Heisenberg group, which is equivalent to a
central extension of commutative space and represents a deformation of commutative space.
The lattice embedding determines how the generators of noncommutative torus
act on the module, representing the translations along the cycles.
With this in mind, we also identify the embedding maps for noncommutative supertori in two dimensions
analyzing the generators of them.

In the super case, we  have to additionally
implement the spin structures due to the presence of fermionic coordinates.
 Odd spin structure is realized by implementing the translation
 in the fermionic direction as in the construction of the noncommutative torus.
Even spin structures are
realized with appropriate versions of the spin angular momentum operator
to express the sign changes of the fermionic coordinates under the translations
along the cycles.\\

\vspace{5mm}
\noindent
{\Large \bf Acknowledgments}

\vspace{5mm} \noindent The authors thank KIAS for hospitality
during the time that this work was done. This work was supported
by the Korea Research Foundation Grants funded by
the Korean Government(MOEHRD),
KRF-2007-313-C00152 (E. C.-Y.),
and KRF-2006-311-C00195 (H. K.),
and is the result of research
activities (Astrophysical Research Center for the Structure and
Evolution of the Cosmos (ARCSEC)) and grant No.\
R01-2006-000-10965-0 from the Basic Research Program supported by
KOSEF (H. N.).
\\




\end{document}